\documentclass{article}
\usepackage{amssymb}
\usepackage{graphicx}
\usepackage{amsmath}
\usepackage{psfrag}

\begin{document}

\title{Constraints on lepton flavor violation in the two Higgs doublet model}
\author{Rodolfo A. Diaz\thanks{%
radiazs@unal.edu.co}, R. Martinez\thanks{%
remartinezm@unal.edu.co}, and Carlos E. Sandoval\thanks{%
cesandovalu@unal.edu.co} \\
Universidad Nacional de Colombia, \\
Departamento de F\'{\i}sica. Bogot\'{a}, Colombia.}
\maketitle

\begin{abstract}
Constraints on the whole spectrum of lepton flavor violating vertices are
shown in the context of the standard two Higgs doublet model. The vertex
involving the $e-\tau $ mixing is much more constrained than the others, and
the decays proportional to such vertex are usually very supressed.


\end{abstract}

\section{Introduction}

Many extensions of the standard model leads to flavor changing neutral
currents (FCNC) naturally. It is the case of models with an extended Higgs
sector. However, owing to the high suppression imposed by experiments,
several mechanisms has been used to get rid of them, such as discrete
symmetries \cite{Glashow}, permutation symmetries \cite{Lin}, and different
textures of Yukawa couplings \cite{Zhou}. Notwithstanding, the increasing
evidence on neutrino oscillations seems to show the existence of mass terms
for the neutrinos as well as of family lepton flavor violation (LFV) \cite%
{Fukuda}. Such fact has inspired the study of many scenarios that predict
LFV processes as in the case of SUSY theories with R- parity broken \cite{R1}%
, SU(5) SUSY models with right-handed neutrinos \cite{Baek}, models with
heavy Majorana neutrinos \cite{Gvetic}, and multi-Higgs doublet models with
right-handed neutrinos for each lepton generation \cite{Lavoura}. On the
other hand, LFV in the charged sector has been also examined in models such
as SUSY GUT \cite{Okada}, and the two Higgs doublet model (2HDM) \cite{Zhou,
KangLee, us}.

Experimental upper limits for the branching ratios of these processes have
been obtained from several collaborations, in the case of the charged lepton
sector, searches for them have been carried out through leptonic and
semileptonic decays of $K$ and $B$ mesons \cite{KB Col}, as well as purely
leptonic processes \cite{Lepton Col}. On the other hand, some collaborations
plan to improve current upper limits of some LFV decays by several orders of
magnitude, by increasing the statistics \cite{Improve bounds}. Other
possible sources of improvement lies on the Fermilab Tevatron and LHC by
means of LFV Higgs boson decays. In particular, Ref. \cite{Cruz} shows that
the flavor changing mode $h\rightarrow \mu \tau $ in the context of MSSM and
of an $E_{6}$-inspired multi-Higgs model with an abelian flavor symmetry;
can be sizable at the CERN-LHC and the Fermilab-Tevatron.

Further, one of the most promising source to look for Higgs mediated flavor
changing neutral currents, lies on the muon colliders. It is because they
have the potentiality to produce Higgs bosons in the $s-$channel, with
substantial rate production at the Higgs mass resonance \cite{Reina}. Some
of the main advantages of muon colliders consists of its negligible
synchrotron radiation and brehmstrahlung, as well as the small beam energy
spread \cite{Reina}. From the theoretical point of view, since Higgs Yukawa
couplings are usually proportional to the lepton mass, they give an
important enhancement to cross sections with Higgs mediated $s-$channels,
respect to the ones in an $e^{+}e^{-}$ collider. The small spread in the
center of mass energy would permit a precision measurements of narrow
resonances, that in turn allow a good determination of the Higgs mass and
Higgs decay width. In particular, for the process $\mu ^{+}\mu
^{-}\rightarrow h\rightarrow \mu \tau $, Ref. \cite{SherCol} have found that
in the context of the 2HDM III, hundreds of such events are expected if $%
m_{h}\leq 140$ GeV with a total integrated luminosity of $1fb^{-1}$ over a
negligible background, providing useful information about the $\mu -\tau $
mixing. Besides, Ref. \cite{SherCol} also found that the process $\mu \mu
\rightarrow e\tau $ could be observable as well, though only some few events
are expected. Notwithstanding, for Higgs boson masses above the $%
h\rightarrow WW^{\ast },ZZ^{\ast }$ threshold ($m_{h}\gtrsim 150$ GeV) the
opening of these new channels decrease dramatically the production of such
FCNC decays.

In a recent previous work \cite{us}, some constraints on LFV have been found
in the framework of the two Higgs doublet model with flavor changing neutral
currents. Specifically, bounds on the vertices $\xi _{\mu \tau },\xi _{e\tau
},\xi _{\mu \mu }$,$\xi _{\tau \tau }$, where obtained based on the $g-2$
muon factor and the leptonic decays $\mu \rightarrow e\gamma ,\tau
\rightarrow \mu \mu \mu $, $\tau \rightarrow \mu \gamma $. Additionally,
upper limits on the decays $\tau \rightarrow e\gamma $ and $\tau \rightarrow
eee$ were estimated, finding them to be highly suppressed respect to the
present experimental sensitivity.

The purpose of this work is to complete the information about the spectrum
of the LFV matrix. With this in mind, we shall use the leptonic processes $%
\tau ^{-}\rightarrow \mu ^{-}\mu ^{-}e^{+},\ \tau ^{-}\rightarrow \mu
^{+}\mu ^{-}e^{-}$ and $\tau ^{-}\rightarrow \mu ^{-}e^{-}e^{+}$ as the
inputs for our constraints.

\section{The decays}

We shall work in the context of the two Higgs doublet model (2HDM) with
flavor changing neutral currents, the so called model type III. The lepton
vertices are described by the following Yukawa Lagrangian%
\begin{eqnarray}
-\pounds _{Y} &=&\overline{E}\left[ \frac{g}{2M_{W}}M_{E}^{diag}\right]
E\left( \cos \alpha H^{0}-\sin \alpha h^{0}\right)  \notag \\
&&+\frac{1}{\sqrt{2}}\overline{E}\xi ^{E}E\left( \sin \alpha H^{0}+\cos
\alpha h^{0}\right)  \notag \\
&&+\overline{\vartheta }\xi ^{E}P_{R}EH^{+}+\frac{i}{\sqrt{2}}\overline{E}%
\xi ^{E}\gamma _{5}EA^{0}+h.c.  \label{Yuk lepton}
\end{eqnarray}%
where $H^{0}$($h^{0}$)$\;$denote the heaviest (lightest) neutral $CP-$even
scalar, and$\;A^{0}\;$is a $CP-$odd scalar. $E\;$refers to the three charged
leptons $E\equiv \left( e,\mu ,\tau \right) ^{T}$ \ and $M_{E},\;\xi _{E}\;$%
are the mass matrix and the LFV matrix respectively, $\alpha \;$is the
mixing angle in the $CP-$even sector. We use the parametrization in which
one of the vacuum expectation value vanishes.

The decays needed to obtain our bounds are given by%
\begin{eqnarray*}
\Gamma \left( \tau ^{-}\rightarrow \mu ^{-}\mu ^{-}e^{+}\right)  &=&\frac{%
m_{\tau }^{5}}{4096\pi ^{3}}\xi _{\mu \tau }^{2}\xi _{e\mu }^{2}\left[
\left( \frac{\sin ^{2}\alpha }{m_{H^{0}}^{2}}+\frac{\cos ^{2}\alpha }{%
m_{h^{0}}^{2}}-\frac{1}{m_{A^{0}}^{2}}\right) ^{2}\right.  \\
&&\left. +\frac{8}{3m_{A^{0}}^{2}}\left( \frac{\sin ^{2}\alpha }{%
m_{H^{0}}^{2}}+\frac{\cos ^{2}\alpha }{m_{h^{0}}^{2}}\right) \right] 
\end{eqnarray*}%
\begin{equation*}
\Gamma \left( \tau ^{-}\rightarrow \mu ^{+}\mu ^{-}e^{-}\right) =\frac{%
m_{\tau }^{5}}{6144\pi ^{3}}\xi _{\mu \tau }^{2}\xi _{e\mu }^{2}\left[
\left( \frac{\sin ^{2}\alpha }{m_{H^{0}}^{2}}+\frac{\cos ^{2}\alpha }{%
m_{h^{0}}^{2}}\right) ^{2}+\frac{1}{m_{A^{0}}^{4}}\right] 
\end{equation*}%
\begin{eqnarray*}
\Gamma \left( \tau ^{-}\rightarrow \mu ^{-}e^{-}e^{+}\right)  &=&\frac{%
m_{\tau }^{5}}{6144\pi ^{3}}\xi _{\mu \tau }^{2}\left\{ \left[ \sin \left(
2\alpha \right) \sqrt{\frac{G_{F}}{\sqrt{2}}}\left( \frac{1}{m_{H^{0}}^{2}}-%
\frac{1}{m_{h^{0}}^{2}}\right) m_{e}\right. \right.  \\
&&\left. \left. +\xi _{ee}\left( \frac{\sin ^{2}\alpha }{m_{H^{0}}^{2}}+%
\frac{\cos ^{2}\alpha }{m_{h^{0}}^{2}}\right) \right] ^{2}+\frac{\xi
_{ee}^{2}}{m_{A^{0}}^{4}}\right\} \ .
\end{eqnarray*}%
Observe that the decays containing two identical particles in the final
state possess interferences involving the pseudoscalar Higgs boson, while
the decays with no identical leptons in the final state do not contain
interference terms with the pseudoscalar. On the other hand, in the
calculation of the decay width $\Gamma \left( \tau ^{-}\rightarrow \mu
^{+}\mu ^{-}e^{-}\right) \ $we neglect the diagrams containing the vertex $%
\xi _{e\tau }$ and keep only the ones proportional to $\xi _{\mu \tau }$, we
make this approximation because previous phenomenological analysis shows a
strong hierarchy between these mixing vertices \cite{us} ($\left\vert \xi
_{e\tau }\right\vert <<\left\vert \xi _{\mu \tau }\right\vert $ by at least
five orders of magnitude).

The corresponding experimental upper limits for these rare processes are 
\cite{dataparticle}%
\begin{eqnarray}
\Gamma \left( \tau ^{-}\rightarrow \mu ^{-}\mu ^{-}e^{+}\right)  &\leq
&3.4\times 10^{-18}\text{\ GeV,}  \notag \\
\Gamma \left( \tau ^{-}\rightarrow \mu ^{+}\mu ^{-}e^{-}\right)  &\leq
&4.07\times 10^{-18}\text{ GeV,}  \notag \\
\Gamma \left( \tau ^{-}\rightarrow \mu ^{-}e^{-}e^{+}\right)  &\leq
&3.85\times 10^{-18}\text{ GeV.}  \label{expbounds}
\end{eqnarray}

\subsection{Bounds on $\protect\xi _{\protect\mu e}$ and $\protect\xi _{ee}$}

In a previous work \cite{us}, the LFV vertices coming from the 2HDM type
III, were constrained by using several pure leptonic processes, the
following bounds for the LFV vertices were found%
\begin{eqnarray}
\xi _{e\tau }^{2} &\lesssim &2.77\times 10^{-14}\ ,  \notag \\
\left\vert \xi _{\mu \mu }\right\vert  &\lesssim &1.3\times 10^{-1}\ , 
\notag \\
7.62\times 10^{-4} &\lesssim &\xi _{\mu \tau }^{2}\lesssim 4.44\times
10^{-2}\ ,  \notag \\
\left\vert \xi _{\tau \tau }\right\vert  &\lesssim &2.2\times 10^{-2}\ .
\label{bounds}
\end{eqnarray}%
Such constraints are valid in most of the region of parameters. Since we
intend to complete the analysis made in \cite{us}, we shall make the same
assumptions which we summarize here for completeness. We settle $%
m_{h^{0}}\approx 115\ $GeV, and $m_{A^{0}}\gtrsim m_{h^{0}}$. In order to
cover a very wide region of parameters, we examine five cases for the
remaining free parameters of the model \cite{us}

\begin{enumerate}
\item When $m_{H^{0}}\simeq 115$ GeV.

\item When $m_{H^{0}}\simeq 300$ GeV and $\alpha =\pi /2$.

\item When $m_{H^{0}}$ is very large and $\alpha =\pi /2$.

\item When $m_{H^{0}}\simeq 300\ $GeV and $\alpha =\pi /4$.

\item When $m_{H^{0}}$ is very large and $\alpha =\pi /4$.
\end{enumerate}

\noindent For all those cases the value of the pseudoscalar mass is swept in the range
of $m_{A^{0}}\gtrsim 115$ GeV.

The vertex $\xi _{\mu e}^{2}$ can be constrained by combining the existing
limits on $\xi _{\mu \tau }^{2}$ given in Eqs. (\ref{bounds}), and the upper
limit on the decay width $\Gamma \left( \tau ^{-}\rightarrow \mu ^{-}\mu
^{-}e^{+}\right) $ given by Eq. (\ref{expbounds}). Alternatively, we can
constrain the same vertex from the decay $\Gamma \left( \tau ^{-}\rightarrow
\mu ^{+}\mu ^{-}e^{-}\right) $. The upper limits on $\xi _{\mu e}^{2}$
obtained from both decays are illustrated in table (\ref{tab:mue}) for the
five cases explained above. 
\begin{table}[tbh]
\centering   
\begin{tabular}{||l||l||l||}
\hline\hline
case & from $\tau ^{-}\rightarrow \mu ^{-}\mu ^{-}e^{+}$ & from $\tau
^{-}\rightarrow \mu ^{+}\mu ^{-}e^{-}$ \\ \hline\hline
1 & $\xi _{\mu e}^{2}\lesssim 5.59\times 10^{-3}$ & $\xi _{\mu
e}^{2}\lesssim 1.0\times 10^{-2}$ \\ \hline\hline
2 & $\xi _{\mu e}^{2}\lesssim 1.5\times 10^{-1}$ & $\xi _{\mu e}^{2}\lesssim
2.7\times 10^{-1}$ \\ \hline\hline
3 & unconstrained & unconstrained \\ \hline\hline
4 & $\xi _{\mu e}^{2}\lesssim 1.35\times 10^{-2}$ & $\xi _{\mu
e}^{2}\lesssim 2.43\times 10^{-2}$ \\ \hline\hline
5 & $\xi _{\mu e}^{2}\lesssim 1.67\times 10^{-2}$ & $\xi _{\mu
e}^{2}\lesssim 3.0\times 10^{-2}$ \\ \hline\hline
\end{tabular}%
\caption{{}Bounds on the mixing vertex $\protect\xi _{\protect\mu e}^{2}$,
based on the processes $\protect\tau ^{-}\rightarrow \protect\mu ^{-}\protect%
\mu ^{-}e^{+}$ and $\protect\tau ^{-}\rightarrow \protect\mu ^{+}\protect\mu %
^{-}e^{-}$ for the five cases cited in the text.}
\label{tab:mue}
\end{table}
We should observe that the upper limits obtained from $\tau ^{-}\rightarrow
\mu ^{+}\mu ^{-}e^{-}$ are less restrictive than the ones coming from $\tau
^{-}\rightarrow \mu ^{-}\mu ^{-}e^{+}$. However, both sets of constraints
lie roughly on the same order of magnitude. From table \ref{tab:mue} we can
extract a quite general bound for the vertex $\xi _{\mu e}^{2}$%
\begin{equation*}
\xi _{\mu e}^{2}\leq 1.5\times 10^{-1}\ ,
\end{equation*}%
valid for most of the region of parameters\footnote{%
We should bear in mind however, that none of the restrictions obtained here,
are valid for the third case explained in the text.}. It worths to say that
other restrictions on this vertex can be gotten from $\mu \rightarrow
e\gamma $ or $\tau \rightarrow e\gamma $ assuming that only the diagrams
with a muon in the loop contribute, instead of the tau as customary.
Notwithstanding, bounds obtained this way are much less restrictive. 
\begin{figure}[tbh]
\psfrag{A}{$m_{A^0}$} \psfrag{ee}{$\xi_{ee}$}
\includegraphics[height=5cm]{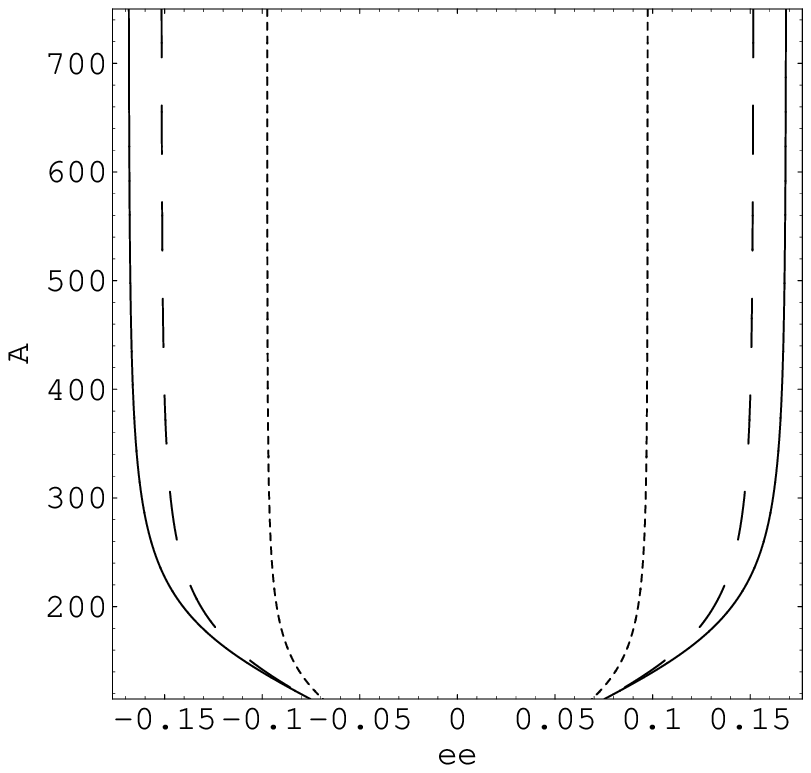} %
\includegraphics[height=5cm]{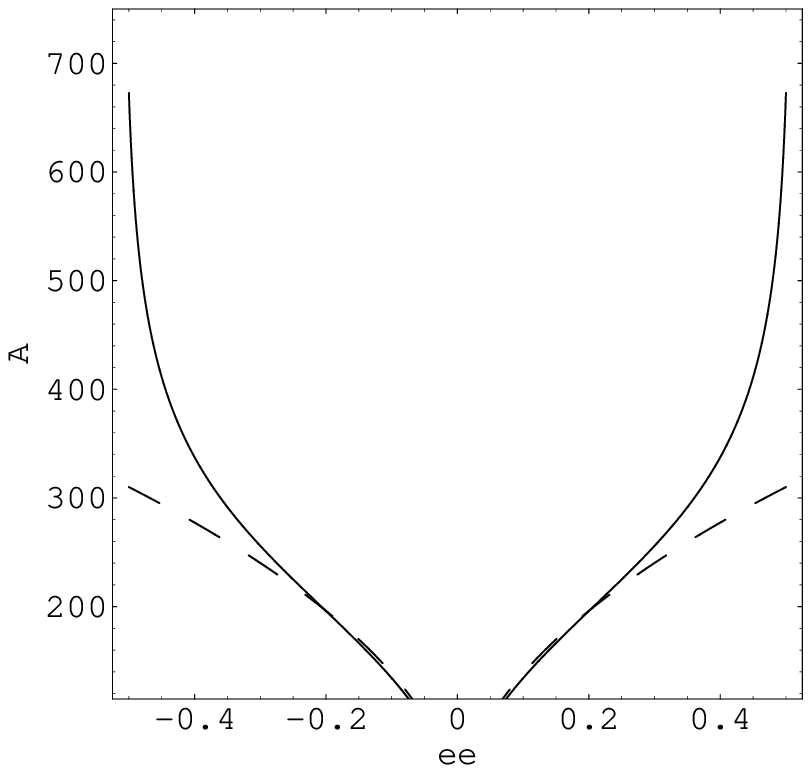}
\caption{Contourplots for the five cases cited in the text in the $\protect%
\xi _{ee}-m_{A^{0}\text{ }}$ plane, based on the process $\protect\tau %
^{-}\rightarrow \protect\mu ^{-}e^{+}e^{-}$. On left: Case 1 (dotted line),
case 4 (dashed line) and case 5 (solid line). On right: Case 2 (solid line)
and case 3 (dashed line).}
\label{fig:contouree}
\end{figure}

On the other hand, we can get contraints on the vertex $\xi _{ee}$ by
combining the already mentioned bounds on $\xi _{\mu \tau }^{2}$ and the
upper experimental constraints for the decay $\Gamma \left( \tau
^{-}\rightarrow \mu ^{-}e^{+}e^{-}\right) $ of Eq. (\ref{expbounds}). Since
the factor $\xi _{ee}$ cannot be factorized in contrast to the case of $\xi
_{\mu e}$, we extract its bounds in the form of contourplots in the $\xi
_{ee}-m_{A^{0}}$ plane, see Fig. (\ref{fig:contouree}). Additionally, we
write in table (\ref{tab:ee}), the constraints obtained for $m_{A^{0}}$ very
heavy and for $m_{A^{0}}\approx 115$ GeV. 
\begin{table}[tbh]
\centering   
\begin{tabular}{||l||l||l||}
\hline\hline
Case & $\left\vert \xi _{ee}\right\vert \left( m_{A^{0}}\text{ very heavy}%
\right) $ & $\left\vert \xi _{ee}\right\vert \left( m_{A^{0}}\text{ }\sim 
\text{115\ GeV}\right) $ \\ \hline\hline
1 & $\lesssim 9.75\times 10^{-2}$ & $\lesssim 6.89\times 10^{-2}$ \\ 
\hline\hline
2 & $\lesssim 5.1\times 10^{-1}$ & $\lesssim 7.41\times 10^{-2}$ \\ 
\hline\hline
3 & unconstrained & unconstrained \\ \hline\hline
4 & $\lesssim 1.5\times 10^{-1}$ & $\lesssim 7.54\times 10^{-2}$ \\ 
\hline\hline
5 & $\lesssim 1.7\times 10^{-1}$ & $\lesssim 7.53\times 10^{-2}$ \\ 
\hline\hline
\end{tabular}%
\caption{{}Bounds for the mixing matrix element $\protect\xi _{ee}$, for $%
m_{A^{0}}\simeq 115$ GeV and for $m_{A^{0}}$ very heavy. Such constraints
are based on the bounds on $\protect\xi _{\protect\mu \protect\tau }$ and
the upper limit for the decay width $\Gamma \left( \protect\tau %
^{-}\rightarrow \protect\mu ^{-}e^{+}e^{-}\right) $.}
\label{tab:ee}
\end{table}
From table (\ref{tab:ee}) we can extract general constraints for $\xi _{ee}$%
, the general bounds read%
\begin{equation*}
\left\vert \xi _{ee}\right\vert \lesssim 5.1\times 10^{-1}\ ;\ \left\vert
\xi _{ee}\right\vert \lesssim 7.54\times 10^{-2}
\end{equation*}%
for $m_{A^{0}}\approx 115$ GeV and for $m_{A^{0}}$ very heavy respectively.
We emphasize again that this prediction is valid in most of the region of
parameters but fails in the case 3 cited above, i.e. when $m_{H^{0}\text{ }}$%
is very large and $\alpha =\pi /2$.

Finally, we make a prediction about the upper limit for the decay width of
the process$\ \tau ^{-}\rightarrow \mu ^{+}e^{-}e^{-}$, based on the limits
on $\xi _{e\mu }$ shown in table \ref{tab:mue} and the limits on $\xi
_{e\tau }$ shown in Eqs. (\ref{bounds}), the results are collected in table %
\ref{tab:taeemu}. Comparing to the present experimental upper limit $\Gamma
\left( \tau ^{-}\rightarrow \mu ^{+}e^{-}e^{-}\right) \leq 3.39\times
10^{-18}\ $GeV the upper limits shown in table (\ref{tab:taeemu}) are at
least ten orders of magnitude smaller (except for the third case). The
strong suppression of this process might be anticipated because of its
proportionality to the vertex $\xi _{e\tau }$ which is much more restricted
than the others \cite{us}.

\begin{table}[tbh]
\centering        
\begin{tabular}{||c||c||}
\hline\hline
Caso & $\Gamma (\tau ^{-}\rightarrow \mu ^{+}e^{-}e^{-})$ \\ \hline\hline
1 & $\lesssim 2.15\times 10^{-29}$GeV \\ \hline\hline
2 & $\lesssim 7.16\times 10^{-29}$ GeV \\ \hline\hline
3 & Unconstrained \\ \hline\hline
4 & $\lesssim 2.91\times 10^{-29}$GeV \\ \hline\hline
5 & $\lesssim 3.21\times 10^{-29}$GeV \\ \hline\hline
\end{tabular}%
\caption{{}Upper limits for the decay width $\Gamma \left( \protect\tau %
^{-}\rightarrow e^{-}e^{-}\protect\mu ^{+}\right) $, based on the contraints
obtained for the LFV vertices $\protect\xi _{\protect\mu e}$ and $\protect%
\xi _{e\protect\tau }$. The experimental upper limit is $3.39\times 10^{-18}$%
GeV}
\label{tab:taeemu}
\end{table}

\section{Conclusions}

\begin{table}[tbp]
\centering        
\begin{tabular}{||l||l||}
\hline\hline
Predictions (in GeV) & Experim. limits \\ \hline\hline
$\Gamma \left( \tau ^{-}\rightarrow e^{-}\gamma \right) \lesssim 1.5\times
10^{-27}$ & $6.12\times 10^{-18}$ \\ \hline\hline
$\Gamma \left( \tau ^{-}\rightarrow e^{+}e^{-}e^{-}\right) \lesssim 5\times
10^{-29}$ & $6.57\times 10^{-18}$ \\ \hline\hline
$\Gamma \left( \tau \rightarrow \mu ^{+}e^{-}e^{-}\right) \lesssim
7.16\times 10^{-29}$ & $3.39\times 10^{-18}$ \\ \hline\hline
\end{tabular}%
\caption{{}Upper limits predicted for some lepton decays. All of them are
highly suppressed respect to the current experimental upper limit.}
\label{tab:pred}
\end{table}
We have found constraints on the whole spectrum of the mixing matrix of
leptons, by using purely leptonic processes. Gathering the information of
all the contraints for LFV vertices, we get the following bounds%
\begin{eqnarray*}
\left\vert \xi _{ee}\right\vert  &\lesssim &5.1\times 10^{-1}\ \ ;\
\left\vert \xi _{\mu e}\right\vert \leq 3.9\times 10^{-1} \\
\left\vert \xi _{e\tau }\right\vert  &\lesssim &1.\allowbreak 66\times
10^{-7}\ ;\left\vert \xi _{\mu \mu }\right\vert \lesssim 1.3\times 10^{-1}\ ;
\\
2.\,\allowbreak 76\times 10^{-2} &\lesssim &\left\vert \xi _{\mu \tau
}\right\vert \lesssim 2.1\times 10^{-1} \\
\left\vert \xi _{\tau \tau }\right\vert  &\lesssim &2.2\times 10^{-2}
\end{eqnarray*}%
In which a strong hierarchy between the vertices $\xi _{\mu \tau }$ and $\xi
_{e\tau }$ is manifest. Additionally, we get predictions for some leptonic
decays shown in table (\ref{tab:pred}), in which we also include the
experimental upper limit for the sake of comparison. From table \ref%
{tab:pred}, we see that such decays are highly suppressed respect to the
current instrumental sensitivity. It owes mainly to its dependence on the $%
\xi _{e\tau }$ vertex which is much more restricted than the others.

We acknowledge the financial support by Fundaci\'{o}n Banco de la Rep\'{u}%
blica.


\begin{thebibliography}{99}
\bibitem{Glashow} S. Glashow and S. Weinberg, Phys. Rev. \textbf{D15}, 1958
(1977).

\bibitem{Lin} C. Lin, C. Lee and Y-W Yang., Chin. J. Phys. \textbf{26,} 180
(1988) ; ibid. Phys. Rev \textbf{D42}, 2355 (1990); arXiv: hep-ph/9311272

\bibitem{Zhou} Marc Sher and Yao Yuan, Phys. Rev. \textbf{D44}, 1461 (1991);
Yu-Feng Zhou, arXiv: hep-ph/0307240.

\bibitem{Fukuda} Y. Fukuda, et. al., Phys. Rev. Lett. \textbf{81}, 1562
(1998).

\bibitem{R1} M. Nowakowski and A. Pilaftsis, Nucl. Phys. \textbf{B461}, 19
(1996); A. Joshipura and M. Nowakowski, Phys. Rev. \textbf{D51}, 5271
(1995); G. Ross and J. W. F. Valle, Phys. Lett. \textbf{B151}, 375 (1985);
A. Kaustubh, M. Graessner Phys. Rev. \textbf{D61}, 075008 (2000).

\bibitem{Baek} S. Baek, T. Goto, Y. Okada and K. Okumura, arXiv:
hep-ph/0109015.

\bibitem{Gvetic} G. Gvetic et. al., Phys. Rev. \textbf{D66}, 034008 (2002).

\bibitem{Lavoura} W. Grimus and L. Lavoura, arXiv: hep-ph/0204070.

\bibitem{Okada} Y. Okada, K. Okumura, Y. Shimizu, Phys. Rev. \textbf{D61},
094001 (2000).

\bibitem{KangLee} S.K. Kang and K.Y.\ Lee, Phys. Lett \textbf{B521},\ 61
(2001).

\bibitem{us} Rodolfo A. Diaz, R. Martinez, and J.-Alexis Rodriguez, Phys.
Rev. \textbf{D67}, 075011 (2003).

\bibitem{KB Col} K. Arisaka, et. al., Phys. Rev. Lett. \textbf{70}, 1049
(1993); K. Arisaka, et. al., Phys. Lett. \textbf{B432}, 230 (1993); A. M.
Lee, et. al., Phys. Rev. Lett. \textbf{64} (1990) 165; G. Lopez Castro, R.
Martinez and J. H. Mu\~{n}oz, Phys. Rev. \textbf{D58}, 033003\textbf{\ }%
(1998); The BABAR Coll., arXiv: hep-ex/0207083; V. Halyo, arXiv:
hep-ex/0207010; CLEO Coll., Phys. Rev. \textbf{D65},\textbf{\ }121801 (2002)
; R. Appel et. al. Phys. Rev. Lett. \textbf{85},  2877 (2000); W. Molzon
Int. J. Mod. Phys. \textbf{A15}S1, 140 (2000) ; D. Pripstein et. al. Phys.
Rev. \textbf{D61}, 032005 (2000).

\bibitem{Lepton Col} M. L. Brooks \emph{et. al.} [MEGA Collaboration], Phys.
Rev. Lett. \textbf{83}, 1521 (1999); P. Wintz, in Proceedings of the \emph{%
First International Symposium on Lepton and Baryon Number Violation}, edited
by H. V. Klapdor-Kleingrothaus and I. V. Krivosheina (Institute of Physics,
Bristol/Philadelphia), p. 534 (1998); C. Dohmen et. al. Phys. Lett. \textbf{%
B317} (1993) 631; U. Bellgardt et. al., Nucl. Phys. \textbf{B299}, 1 (1998);
R. D. Bolton, \emph{et. al}., Phys. Rev. \textbf{D38}, 2077 (1983); M. L.
Brooks, \emph{et. al.}, Phys. Rev. Lett. \textbf{83}, 1521 (1999); Y. Okada,
K. Okumura, Y. Shimizu, Phys. Rev. \textbf{D61}, 094001 (2000).

\bibitem{Improve bounds} L. M. Barkov \emph{et. al.}, research proposal to
PSI; S. Ritt, in Proceedings of \emph{The 2nd International Workshop on
Neutrino Oscillations and their Origin, }edited by Y. Suzuki \emph{et.al.\ }%
(World Scientific), p. 245 (2000). M. Bachman \emph{et. al.,} [MECO
Collaboration], experimental proposal E940 to BNL AGS (1997). Y. Kuno, in
proceedings of the \emph{2nd International Workshop on Neutrino Oscillations
and their Origin}, edited by Y. Suzuki \emph{et. al.} (World Scientific), p.
253 (2000).

\bibitem{Cruz} J. Lorenzo Diaz-Cruz, arXiv: hep-ph/0207030; Chung Kao,
arXiv: hep-ph/0210001.

\bibitem{Reina} D. Atwood, L. Reina and A. Soni, Phys. Rev. D \textbf{55},
3156 (1997); Phys. Rev. Lett. \textbf{75}, 3800 (1995); E. Asakawa, A.
Sugamoto, and I. Watanabe [arXiv: hep-ph/0004005]; B. Grzadkowski, and J.
Pliszka [arXiv: hep-ph/0004034]; R. Casalbuoni [arXiv: hep-ph/0105091]; R.
A. Alanakyan [arXiv:hep-ph/0107092]; M. S. Berger [arXiv: hep-ph/0110390];
H. Fraas \emph{et. al.} [arXiv: 0303044].

\bibitem{SherCol} Marc Sher, Phys. Lett. \textbf{B487}, 151 (2000).

\bibitem{dataparticle} Particle Data Group, K. Hagiwara \emph{et. al.} ,
Phys. Rev. \textbf{D66,} 010001 (2002) and references therein.
\end{thebibliography}
\end{document}